\documentstyle[aps]{revtex}
\begin{document}
\draft
\title{\bf Condensation of a Hard-Core Bose Gas}
\author{K. Ziegler}
\address{Institut f\"ur Physik, Universit\"at Augsburg\\
D-86135 Augsburg, Germany}
\date{\today}
\maketitle
\begin{abstract}
A grand canonical system of hard-core bosons, subject to thermal
fluctuations, is studied on a
lattice. Starting from the slave-boson representation with fields
for occupied and unoccupied sites, an effective field theory is
derived in which a complex field corresponds with the order parameter
of the condensate and a real field with the total density of bosons.
Near the boundary between the normal and the superfluid phase we obtain
the Ginzburg-Landau functional for the superfluid order parameter. 
A mean-field calculation shows that the
critical temperature $T_c$ increases with increasing density
up to a maximum and decreases with further increasing density.
\end{abstract}

\section{Introduction}

\noindent
Interacting bosonic quantum systems are of special interest 
because of the effect of Bose condensation. The classical
example is the Bose-Einstein condensation in an ideal
Bose gas \cite{bose,einstein}. The analogous phenomenon in a real (i.e. 
interacting) Bose gas
is the transition from a normal to a superfluid state (e.g., in $^4$He). 
Another related phenomenon is the condensation of a cold gas of
bosonic atoms in a magnetic trap \cite{anderson}.
There is a number of recent theoretical investigations of the effect
of the interaction on the critical properties of the
normal-superfluid transition based on Monte Carlo simulations
\cite{grueter,holzmann} and analytic calculations \cite{baym,baym1}.

\noindent
A fundamental model for the description of interacting bosons is the
Ginzburg-Landau theory, usually called the Gross-Pitaevskii
approach to Bose condensates \cite{ginzburg,gross}. However,
there are dense systems of bosons (like the classical $^4$He 
superfluid), for which the Ginzburg-Landau theory is valid only very 
close to the phase transition from the normal fluid to the superfluid,
where the order parameter is small. Away from the phase
transition the interaction of the order parameter field
is more complicated than described by this theory. Moreover, the
Ginzburg-Landau theory is only a description of the order
parameter (which is related to the density of the superfluid)
but does not take into account the interaction with the non-superfluid part
of the system. Consequently,
it does not provide reliable information, e.g., for the value of the critical
temperature. The question of the effect of the interaction on the
latter has been discussed in the recent literature in
great detail and with controversal results \cite{grueter}. It seems that
the critical temperature can be shifted by a variation of the
density of bosons. In particular, at low density it was found
that the shift $\Delta T_c/T_0\sim(na^3)^\alpha$, where $T_0$
is the critical temperature of the ideal (non-interacting) Bose gas,
$n$ the total density of bosons and $a$ the scattering
radius of the hard-core interaction. Depending on the
calculational method and approximations, the exponent $\alpha$ varies
between $1/3$ \cite{stoof} and $1/2$ \cite{huang}. Recent Monte Carlo 
simulations \cite{grueter,holzmann} support
$\alpha=1/3$. This value was also obtained by a self-consistent calculation
of the quasiparticle spectrum \cite{baym} and in a $1/N$ expansion 
\cite{baym1}. At a high density the critical temperature reaches
a maximum and decreases with even higher densities. The latter is
a consequence of the depletion of the condensate due to interaction.

In order to give a complete overview of the
properties of an interacting Bose gas we need a model which takes
fully into account all parts of the system of bosons. Close to the
critical point, however, it should lead to the Ginzburg-Landau theory.
Such a model was given by hard-core
bosons, based on a slave-boson representation \cite{ziegler1}. Here
we will briefly discuss this model and evaluate its critical temperature
for different densities.

The paper is organized as follows: In Sect. 2 the slave-boson respresentation
of hard-core bosons on a lattice is introduced. Then in Sect. 3 two
collective fields are defined. One represents the superfluid condensate the 
other the total density of bosons, as discussed in Sect. 4. In
Sect. 5 the total density, the condensate density and the critical
temperature are calculated in mean-field approximation. Finally, concluding
remarks and a discussion are given in Sect. 6.

\section{The Model: Slave-Boson Representation}

\noindent
A continuous system of hard-core bosons with scattering length $a$ is
approximated by a lattice Bose gas with lattice constant $a$. Although
this approximation is limited because it restricts configurations of bosons
to be commensurate with the lattice structure, it is more suitable for
the investigation of a dense system of bosons than the usually considered
$|\Phi|^4$ (Ginzburg-Landau) theory. The representation of the
model uses the slave-boson approach to hard-core bosons \cite{ziegler1}.
(Originally the slave-boson approach was invented for the (fermionic) Hubbard
model \cite{barnes,kotliar}.) 
The latter relies on a picture in which a particle
trades its position with an empty site on the lattice. Both, the particle
as well as the empty site, are described by corresponding creation and
annihilation operators. In a
functional integral representation of a grand canonical ensemble of
bosons with chemical potential $\mu$ at temperature $T$
this can be formulated in terms of a complex field 
$b_x$ of bosons and a complex field $e_x$ of empty sites with the action
$$
S_{s.b.}={1\over T}\sum_{x,x'}b^*_xe_xt_{x,x'}b_{x'}e^*_{x'}
-{1\over T}\sum_x\mu |b_x|^2,
$$
where the first term describes the exchange of bosons and empty sites
in a hopping process at sites $x$ and $x'$ with rate $t_{x,x'}$.
A local constraint $|e_x|^2+|b_x|^2=1$ takes care of the
complementary character of the bosons and empty sites. In $S_{s.b.}$
we consider only the thermal fluctuations (i.e. a vanishing Matsubara
frequency) because non-zero Matsubara frequencies are separated by
a gap if the temperature $T$ is non-zero. Here we assume that the temperature
enters through the action $S_{s.b.}$ but not through the constraint.
Physical quantities can be calculated from the partition function
$$
Z=\int e^{-S_{s.b.}}\prod_x\delta(|e_x|^2+|b_x|^2-1)
db_xdb^*_xde_xde^*_x.
$$
For instance, the total density of bosons $n$ is given as the
response to a change of the chemical potential $\mu$
\begin{equation}
n={T\over N}{\partial\log Z\over\partial\mu}
={T\over N}{1\over Z}{\partial Z\over\partial\mu},
\label{density}
\end{equation}
where $N$ is the number of lattice sites.
To evaluate the density of the condensate $\rho_s$ we must add an external
vector potential to the action $S_{s.b.}$ (i.e. a Peierls factor to the
hopping matrix $t$) and measure the response to this potential
\cite{ziegler2}. We
shall return to the corresponding expression subsequently.


\section{Collective-Field Representation}

Since the fields $b_x$ and $e_x$ are subject to the local constraint, we
can not treat them in a conventional way as order parameter fields. It is
necessary to eliminate the constraint in the partition function which
can be achieved by integration over the fields.
For this purpose we introduce a complex collective field $\Phi_x$ and a real
field $\varphi_x$ which break up the biquadratic term in the action
$S_{s.b.}$. This can be written as the (Hubbard-Stratonovich) transformation
\begin{equation}
S_{s.b.}\to S'=T\sum_{x,x'}\Phi_x(1-t)^{-1}_{x,x'}\Phi^*_{x'}+T\sum_x
\varphi_x^2+\sum_x\pmatrix{
e_x\cr
b_x\cr
}\cdot\pmatrix{
2\varphi_x+T^{-1}&\Phi_x\cr
\Phi^*_x&-\mu T^{-1}\cr
}\pmatrix{
e^*_x\cr
b^*_x\cr
}.
\label{coll}
\end{equation}
The new action $S'$ gives the same partition function which can be
seen by integrating over the collective fields. The field $\varphi$ is
necessary in order to invert the hopping term
which is now the positive matrix $1-t$. Then we can perform the
integration of the slave-boson fields. This becomes simple if the
$2\times2$ Hermitean matrix in Eq. (\ref{coll}) is diagonalized by a unitary
transformation. The latter leaves the constraint $|b_x|^2+|e_x|^2=1$
invariant and gives the eigenvalues
$$
\lambda_{x,\pm}=\varphi_x+1/2T-\mu/2T
\pm\sqrt{(\varphi_x+1/2T+\mu/2T)^2+|\Phi_x|^2}.
$$
Now the partition function reads
$$
Z=Z_0\int e^{-S_b}\prod_x\delta(|b_x|^2+|e_x|^2-1)
e^{-T\varphi_x^2-\lambda_{x,+}|e_x|^2-\lambda_{x,-}|b_x|^2}
$$
$$
\times db_xdb^*_xde_xde^*_xd\varphi_xd\Phi_xd\Phi^*_x,
$$
where the only non-local term is
$$
S_b=T\sum_{x,x'}\Phi_x(1-t)^{-1}_{x,x'}\Phi^*_{x'}.
$$
The $e_x$ and $b_x$ integration can be carried out (s. Appendix A)
which yields
$$
Z=Z_0\int e^{-S_b}\prod_xe^{-T\varphi_x^2}
{e^{-\lambda_{x,+}}-e^{-\lambda_{x,-}}
\over\lambda_{x,+}-\lambda_{x,-}}d\varphi_xd\Phi_xd\Phi^*_x
$$
$$
=Z_0\int e^{-S_b}\prod_xe^{-1/4T+\mu/2T}\int e^{-T\varphi_x^2}{sinh(
\sqrt{(\varphi_x+\mu/2T)^2+|\Phi_x|^2})\over\sqrt{(\varphi_x+\mu/2)^2
+|\Phi_x|^2}}d\varphi_xd\Phi_xd\Phi^*_x.
$$
The constant $Z_0$ is the normalization factor of the 
Hubbard-Stratonovich transformation of Eq. (\ref{coll}). It is convenient to
separate the field-independent factor $\prod_xe^{-1/4T+\mu/2T}$ to define
the partition function
$${\bar Z}
=\int e^{-S_b}\prod_x\int e^{-T\varphi_x^2}{sinh(
\sqrt{(\varphi_x+\mu/2T)^2+|\Phi_x|^2})\over
\sqrt{(\varphi_x+\mu/2T)^2+|\Phi_x|^2}}
d\varphi_xd\Phi_xd\Phi^*_x.
$$
This partition function has the effective action for
the collective field $\Phi_x$
$$
S=S_b+S_0,
$$
where $S_b$ is its non-local (i.e. hopping) part and
$$
S_0=\sum_x\log(Z_1(|\Phi_x|^2)
$$
is its local (i.e. potential) part with
$$
Z_1(|\Phi|^2)
= \int e^{-T\varphi^2}{sinh(\sqrt{(\varphi+\mu/2T)^2+|\Phi|^2})
\over\sqrt{(\varphi+\mu/2T)^2+|\Phi|^2}}d\varphi.
$$

\section{Interpretation of the Fields}

The introduction of the collective field has completely separated the
hopping
part $S_b$ from the potential part $S_0$. The latter does not depend on the
phase of the collective field. The hopping part alone describes free bosons
(ideal Bose gas) with the usual complex field. This can be seen by
writing the partition function of the ideal Bose gas as
$$
Z_{IBG}=\prod_k\Big[\sum_{n_k\ge0}e^{- n_k(\epsilon_k-\mu_0)/T}\Big]
=\prod_k[1-e^{-(\epsilon_k-\mu_0)/T}]^{-1},
$$
where $k$ is a quantum number which characterizes the system of bosons.
This can also be expressed in terms of an integral over a complex field as
$$Z_{IBG}=\int\exp\Big(-\sum_k[1-e^{-(\epsilon_k-\mu_0)/T}]
|\Phi_k|^2\Big)\prod_kd\Phi_kd\Phi^*_k/\pi.
$$
(Notice that a rescaling of the field yields only a factor to $Z_{IBG}$.)
For the translational-invariant Bose gas $k$ is the wavevector and we have
the energy
$$
\epsilon_k={\hbar^2k^2\over2m}
$$
with the boson mass $m$.

Using the Fourier components $(1-{\tilde t}_k)^{-1}$ in the non-local
term $S_b$ we can compare the latter with the corresponding expression of
the ideal Bose gas 
\begin{equation}
\sum_k[1-e^{-(\epsilon_k-\mu_0)/T}]|\Phi_k|^2.
\label{ideal}
\end{equation}
By setting the hopping term $(1-{\tilde t}_k)^{-1}$
equal to $1-e^{-(1/T)(\epsilon_k-\mu_0)}$ we obtain
$$
{\tilde t}_k=(1-e^{(\epsilon_k-\mu_0)/T})^{-1}.
$$
The chemical potential of the ideal Bose gas is restricted to $\mu_0\le0$
whereas it can have any real value in the interacting Bose gas.
In particular, we can choose $\mu_0\ge0$ such that ${\tilde t}_k>0$.
Moreover, in the Bose gas we can now apply the continuum
approximation
\begin{equation}
S_b/T=\sum_{x,x'}\Phi_x(1-t)^{-1}_{x,x'}\Phi^*_{x'}
\approx\int\Big[-{\hbar^2\over2m}\Phi_x(\nabla^2\Phi^*)_{x}
+\alpha |\Phi_x|^2\Big]d^3x,
\label{dilute}
\end{equation}
where $\alpha=\sum_x(1-t)^{-1}_{x,x'}$.
It is important to notice that $\alpha=(1-{\tilde t}_0)^{-1}$ is
positive because $1-t$ was defined as a positive matrix. This implies
that $S_b$ is a positive quadratic form.
$S_b$ can be compared with the expression of the ideal Bose gas if we
replace $\alpha$ by $-\mu_0$. Then the expression in (\ref{dilute}) is an
approximation of Eq. (\ref{ideal}) for small $\epsilon_k-\mu_0$,
which applies, e.g., to high temperatures. This approximation
is very common in the literature \cite{fetter,popov} and can also be used in
the case of hard-core bosons. 

Physical quantities can be expressed as expectation values of the new
fields $\varphi$ and $\Phi$. For instance, we obtain from Eq. (\ref{density})
for the total density of bosons
the expression (see Appendix B)
$$
n={1\over2}+T\langle\varphi_x\rangle,
$$
where
\begin{equation}
\langle...\rangle={1\over{\bar Z}}
\int ...\ e^{-S_b}
\prod_xe^{-T\varphi_x^2}{sinh(\sqrt{(\varphi_x+\mu/2T)^2+|\Phi_x|^2})
\over\sqrt{(\varphi_x+\mu/2T)^2+|\Phi_x|^2}}d\varphi_x\Phi_xd\Phi^*_x.
\label{density2}
\end{equation}
Thus $\varphi_x$, which is conjugate to $|e_x|^2$ according to Eq. 
(\ref{coll}),
is related to the total density of bosons.
Conversely, $\Phi_x$, which is conjugate to $e_xb_x^*$ according to 
Eq. (\ref{coll}), corresponds to the density of the condensate
\cite{ziegler1,ziegler2} and can be expressed as the expectation value
$$
\rho_s=T\alpha\langle|\Phi_x|^2\rangle
={T\alpha\over{\bar Z}}\int|\Phi_x|^2e^{-S_b-S_0}\prod_xd\Phi_xd\Phi^*_x.
$$
We notice that the potential part of the partition function $S_0$ is
symmetric with respect to $\mu\to-\mu$. This implies that the 
expectation value $\langle\varphi_x\rangle$ is an {\it odd} function of $\mu$.
Therefore, the density varies monotoneously with $\mu$, as it should.
The symmetry of $S_0$ with respect to $\mu\to-\mu$ implies that
$\rho_s$ is an {\it even} function of $\mu$. This is a characteristic
feature of our lattice hard-core bosons in which bosons and empty
sites are dual to each other. 

\subsection{Near the Critical Point}

The $\varphi$-integration in $Z_1$ can be performed numerically in order to
obtain an effective potential 
$$
-\sum_x\log(Z_1(|\Phi_x|^2).
$$
It is interesting to notice that the expression $\log(Z_1(|\Phi_x|^2)$ is 
linear for large values of $|\Phi_x|$ (cf. Fig. 1).
Therefore, the quadratic term in $S_b$
suppresses the large fluctuations. This means that the $|\Phi_x|^4$
approximation of the Ginzburg-Landau theory provides a stronger suppression
of large fluctuations than the complete theory.

\begin{figure}
\begin{center}
\input{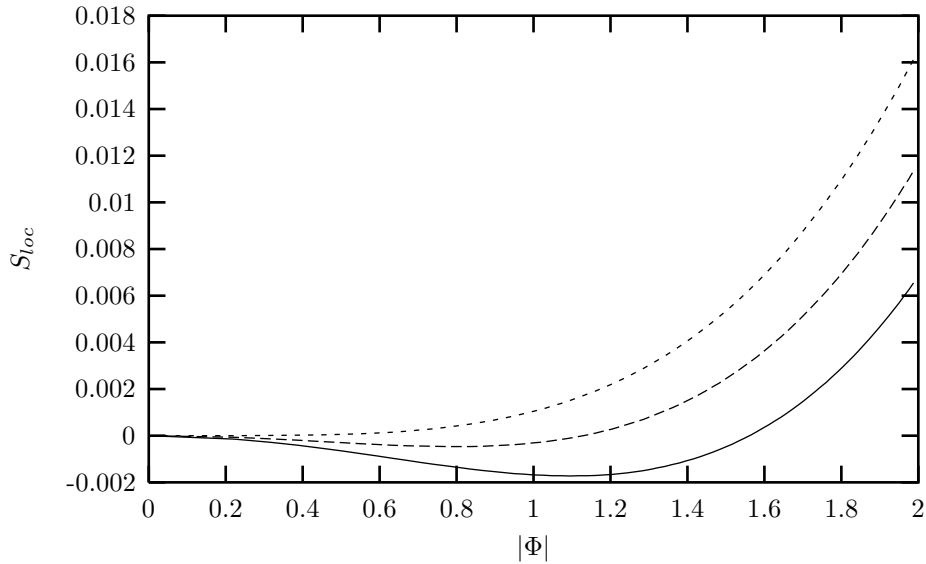}
\end{center}
\caption{
Behavior of the local part of the action $S_{loc}=\alpha T |\Phi|^2
-\log(Z_1(|\Phi|^2))$ for $T=0.1$, $\alpha=1.02$ and three different
values $\mu=0,0.07,0.1$. There is a critical point at $\mu\approx0$.}
\end{figure}

Near the critical point we can expand the free energy around $|\Phi_x|=0$
because the order parameter field $\Phi$ is small. The result of this
expansion is the Ginzburg-Landau functional for the collective field with
$$
T\sum_{x,x'}\Phi_x(1-t)^{-1}_{x,x'}\Phi^*_{x'}
-\sum_x\Big(b_1|\Phi_x|^2+b_2|\Phi_x|^4\Big)
$$
\begin{equation}
\approx\int\Big[
-{T\hbar^2\over2m}\Phi_x(\nabla^2\Phi^*)_{x}
+(T\alpha-b_1)|\Phi_x|^2-b_2|\Phi_x|^4
\Big]d^3x,
\label{GLE}
\end{equation}
where
$$
b_1={Z_1'(0)\over Z_1(0)},\ \ \ 
b_2={1\over2}\Big[{Z_1''(0)\over Z_1(0)}
-\Big({Z_1'(0)\over Z_1(0)}\Big)^2\Big].
$$
Expression (\ref{GLE}) is also known as the Gross-Pitaevskii functional
which is often used as a model of a dilute Bose system \cite{ginzburg,gross}.
In our derivation we have $\mu$- and $T$-dependent coefficients.
As one can see in Fig. 1 the term with $|\Phi|^2$ can change its sign
with a changing $\mu$. On the other hand, away from $\Phi=0$ the
behavior is controlled by the confining part $|\Phi|^4$. This is the
typical Ginzburg-Landau picture of a second-order phase transition.

The effect of the $|\Phi|^4$ interaction on $T_c$ in the model of
Eq. (\ref{GLE}) was 
recently investigated in detail in a self-consistent calculation \cite{baym}
and in a $1/N$ expansion \cite{baym1}. 
In this model it is assumed that the coefficient of the $|\Phi|^4$ term
depends on the the scattering length $a$ and the total density by a factor
$a^3n$. This is valid at low density. The calculation
shows that the critical temperature of the interacting system $T_c$,
normalized with the critical temperature of the ideal Bose gas,
is shifted as
\begin{equation}
1-T_0/T_c \sim c_0 a n^{1/3}.
\label{shift}
\end{equation}
\begin{table}
\caption{
The coefficients of the critical temperature shift in Eq. (\ref{shift}) from 
different works}
\begin{tabular}{c c c c c}
Reference: & \cite{grueter} & \cite{holzmann} &\cite{baym} & \cite{baym1}\\
\tableline
$c_0$: & 0.34 & 2.2 & 1.5 & 2.33 \\
\end{tabular}
\end{table}
\vfill
\noindent
Numerical investigations \cite{grueter} show that the critical temperature
must descrease at higher densities. This depletion effect, which cannot
be seen in the Gross-Pitaevskii approach, will be discussed
in the slave-boson approach subsequently.

\section{Mean-Field Approximation}

Since $\varphi_x$ is a field which appears only in local terms, it can be
integrated out at each site independently for a given value of the
condensate field $\Phi_x$. The treatment of $\Phi_x$ is more difficult
since it appears in non-local term $S_b$. It can be studied in terms of the
classical field equation
\[
\Big[-{\hbar^2T\over2m}\nabla^2
+T\alpha-{Z_1'(|\Phi_x|^2)\over Z_1(|\Phi_x|^2)}\Big]\Phi_x=0,
\]
which is the extremum of the action $S_b+S_0$. A further simplification
is the additional assumption that the condensate field varies only
weakly in space: $\nabla^2\Phi\approx0$. This gives the mean-field or
Thomas-Fermi approximation.
Both densities, $n$ and $\rho_s$, can be evaluated in mean-field
approximation. The mean-field free energy reads
\begin{equation}
F_{MF}=-{1\over N}\log Z ={1\over4}-{\mu\over2T}
+T\alpha |\Phi|^2-\log(Z_1(|\Phi|^2)),
\label{freee}
\end{equation}
where $|\Phi|$ must be at the minimum of $F_{MF}$. 
For a given chemical potential $\mu$ there is a critical value
$T_c$ which separates two regimes: one regime for
$T<T_c$ in which the minimum of the mean-field free energy is
$|\Phi|^2>0$ and another regime with $|\Phi|^2=0$ for
$T\ge T_c$.
This can be seen by plotting $T\alpha|\Phi|^2-\log(Z_1(|\Phi|^2))$
(Fig. 1)
which has, depending on $\mu$, either a minimum at $|\Phi|=0$ or another
one at $|\Phi|\ne0$. \cite{ziegler1}
The minimal value $|\Phi|$ varies continuously as one goes
through $T_c\alpha$. The behavior of the densities as functions of
$\mu$ at a fixed temperature is shown in Fig. 2.

\begin{figure}
\begin{center}
\input{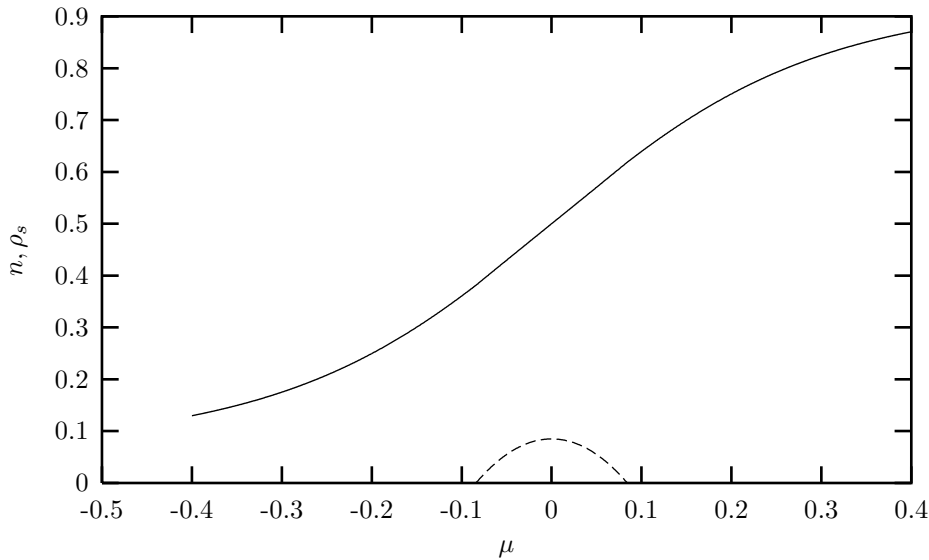}
\end{center}
\caption{Mean-field result of the total density $n$ (full curve) and the
superfluid density $\rho_s$ (dashed curve) at $T\alpha=0.1$.}
\end{figure}

\subsection{Near the Critical Point}

For a very dilute system the mean-field approximation is insufficient
and the calculations of Refs. \cite{baym,baym1} should be applied.
However, at higher densities (more than $n\approx 0.2$) the
mean-field approximation should be reliable.
Then the critical temperature $T_c$ of the mean-field calculation is
$$
T_c=b_1/\alpha\equiv Z_1'(0)/(\alpha Z_1(0)).
$$
The decreasing behavior of $T_c/T_0$ ($T_0\propto n^{2/3}$ is the condensation
temperature of the ideal Bose gas) is shown in Fig. 3. Our mean-field
result for $T_c/T_0$ agrees qualitatively with the Monte Carlo result
of Ref. \cite{grueter}. However, we expect that fluctuations might reduce
the critical temperature substantially close to $n=1$.

\begin{figure}
\begin{center}
\input{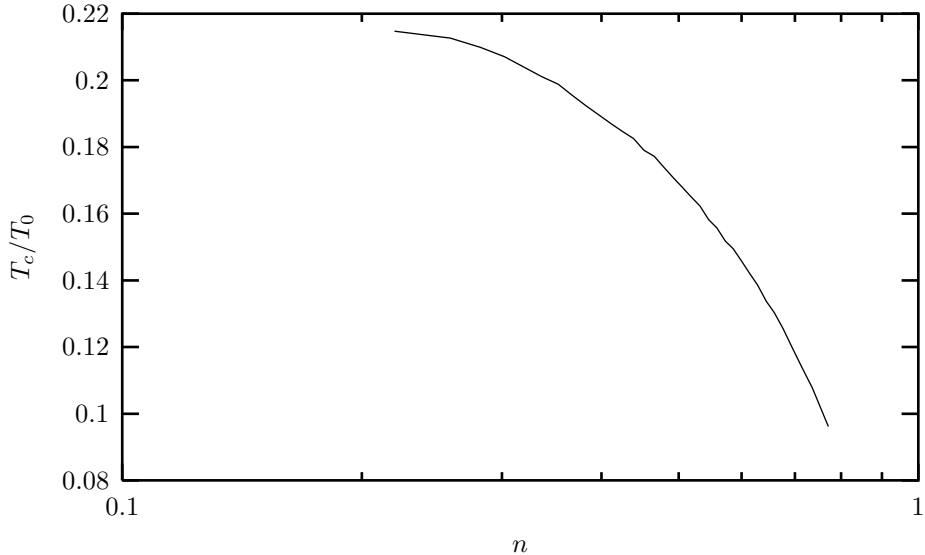}
\end{center}
\caption{Mean-field result of the critical temperature $T_c$ of the
hard-core bosons in arbitrary units, normalized with the critical
temperature of the ideal bose gas $T_0\propto n^{2/3}$. This behavior is
similar to the result found in a Monte Carlo simulation of 
Ref. [1]}
\end{figure}

We can expand the mean-field free energy (\ref{freee}) in powers of
$|\Phi|^2$ up to $|\Phi|^4$. Then the minimum of $F$ must satisfy
the mean-field equation
\begin{equation}
|\Phi|^2\sim -{T\alpha Z_1(0)-Z_1'(0)\over T\alpha Z_1'(0)-Z_1''(0)}
={T_c-T\over(T-T_c)b_1-2b_2/\alpha}
\sim{T_c-T\over-2b_2/\alpha}\ \ \ (T\sim T_c)
\label{mf}
\end{equation}
if the right-hand side is non-negative and $|\Phi|^2=0$ otherwise.
Since $b_2\le 0$ in a typical situation (cf. Fig. 1), there is a non-zero
solution.
The coefficient on the right-hand side of (\ref{mf}) can be evaluated
numerically for a given value of $\mu$. 

\section{Discussion}

The slave-boson representation is given by two fields $b_x$, $e_x$ which are
subject to a constraint: one represents empty the other singly occupied 
sites. These two fields are replaced by two collective fields $\varphi_x$ 
and $\Phi_x$ which have a direct physical interpretation. The former couples 
to $|e_x|^2$ and the latter couples to the product $e_xb_x^*$.
The introduction of the 
collective field $\Phi_x$ enables us to integrate the slave-boson fields
and  to eliminate the constraint of the these fields. 
The effective field theory provides a ``two-liquid'' theory which is 
represented
by the fields $\varphi_x$ and $\Phi_x$. These fields correspond with
the total density $n$ and and the superfluid density $\rho_s$,
respectively. The
latter is the order parameter of the condensation whereas the total
density $n$ does not indicate a critical behavior at the condensation
point (cf. Fig. 2). Therefore, $n$ can be fixed near the critical point
and the theory can be expanded in terms of a small order parameter
field $\Phi_x$. This yields the well-known Ginzburg-Landau
(or Gross-Pitaevskii) theory with density-dependent parameters.
In other words, there is a phase boundary in the $\mu$-$T$ phase
diagram which separates the normal from the superfluid phase. In the
vicinity of this phase boundary we can apply the Gross-Pitaevskii
approach.
 
The condensate field can be treated in mean-field
approximation, assuming a homogeneous order parameter. This reveals
the phenomenon of depletion of the condensate due to a strong interaction
among the bosons which has been observed in superfluid $^4$He. In the
slave-boson representation it
is a consequence of the duality of empty and singly occupied lattice sites
which is reflected by the constraint $|e_x|^2+|b_x|^2=1$. Thus physical
quantities are symmetric with respect to a half-filled system (i.e. 
$n=1/2$ or $\mu=0$).

The lattice theory is not very accurate at densities $n\approx 0.5$
because it reduces the motion of bosons significantly in comparison
with a continuous system. Moreover, the mean-field approximation neglects
vortices and flucuations of the order
parameter. From this point of view we can only expect a qualitative
agreement of our results with those, e.g., obtained from experiments.
Nevertheless, the mean-field approximation should be reasonable if the
density is not too low. The critical exponent of the order parameter
should be renormalized due to fluctuations, using the renormalization
group for the three-dimensional $|\Phi|^4$ model.

The main result is that the phase diagram of the slave-boson theory 
of the strongly interacting bose system
has two normal phases (i.e. a dense and a dilute one) and a superfluid
phase for intermediate densities.
Near the transition points a Gross-Pitaevskii approach can be used
to describe the physics of small order parameter field.
The superfluid density is low in our hard-core system. This might be a
consequence of the strong interaction which suppresses the superfluid
density.


In conclusion, we established an effective field theory which enables
us to evaluate the properties of a strongly interacting system of
bosons. It takes into account the
order parameter of the condensate and the total density by interacting
fields. It describes the phase transition between a normal phase and a 
condensed phase. The phase transition were studied in mean-field 
approximation. We evaluated the density-dependent critical temperature
at densities $n>0.2$ in which the mean-field approximation of the order
parameter is reliable.
\\

\noindent
Acknowledgement:

\noindent
I am grateful to S. Girvin for bringing Ref. [1] to my attention.
\vfill
\eject

\vfill
\eject
\centerline{\bf Appendix A}

\noindent
The integration of the $e$- and the $b$-field can be performed in $Z$
for each point $x$ independently
$$
\int \delta(|b_x|^2+|e_x|^2-1)
e^{-\lambda_{x,+}|e_x|^2-\lambda_{x,-}|b_x|^2}
db_xdb^*_xde_xde^*_x. \eqno (A.1)
$$
The integrand does not depend on the phases of the field. Therefore,
the phase integration contributes a factor $4\pi^2$. Moreover, we
set $s:=|b_x|^2$ and $t:=|e_x|^2$ which yields for (A.1)
$$
\pi^2\int_0^\infty\int_0^\infty\delta(s+t-1)
e^{-\lambda_{x,+}t-\lambda_{x,-}s}dsdt
=\pi^2\int_0^1e^{-\lambda_{x,+}t-\lambda_{x,-}(1-t)}dt
$$
$$
=-\pi^2{e^{-\lambda_{x,+}}-e^{-\lambda_{x,-}}
\over\lambda_{x,+}-\lambda_{x,-}}
$$
\\

\centerline{\bf Appendix B}

\noindent
To write the total density of bosons we have to evaluate
$${T\over N}{\partial\log Z\over\partial\mu}=
{1\over2}
+{T\over N}{1\over{\bar Z}}{\partial{\bar Z}\over\partial\mu}.
$$
Differentation yields
$$
{\partial{\bar Z}\over\partial\mu}
=\int\sum_{\bar x}{\partial Z_1(|\Phi_{\bar x}|^2)\over\partial\mu}
\Big[\prod_{x\ne{\bar x}}Z_1(|\Phi_x|^2)\Big]
\prod_xd\Phi_xd\Phi^*_x.
$$
Since
$$
{\partial Z_1(|\Phi_{\bar x}|^2)\over\partial\mu}
=\int \varphi e^{-T\varphi^2}{sinh(\sqrt{(\varphi+\mu/2T)^2
+|\Phi_{\bar x}|^2})\over\sqrt{(\varphi+\mu/2T)^2+|\Phi_{\bar x}|^2}}
d\varphi
$$
we can write for the previous expression
$$
{\partial{\bar Z}\over\partial\mu}
=\int\sum_{\bar x}\varphi_{\bar x}e^{-S_b}\prod_x
e^{-T\varphi^2}{sinh(\sqrt{(\varphi+\mu/2T)^2+|\Phi_{\bar x}|^2})
\over\sqrt{(\varphi+\mu/2T)^2+|\Phi_{\bar x}|^2}}
d\varphi_xd\Phi_xd\Phi^*_x.
$$
This implies
$$
{1\over{\bar Z}}{\partial{\bar Z}\over\partial\mu}
={1\over{\bar Z}}\sum_{\bar x}\int\varphi_{\bar x}e^{-S_b}
\int e^{-T\varphi_{{\bar x}}^2}{sinh(\sqrt{(\varphi_{{\bar x}}+\mu/2T)^2
+|\Phi_{\bar x}|^2})
\over\sqrt{(\varphi_{{\bar x}}+\mu/2T)^2+|\Phi_{\bar x}|^2}}
d\varphi_{{\bar x}}
\Big[\prod_{x\ne{\bar x}}
Z_1(|\Phi_x|^2)\Big]
\prod_x d\Phi_xd\Phi^*_x
$$
$$
=\sum_{\bar x}\langle\varphi_{\bar x}\rangle.
$$

\end{document}